\documentclass[document,twocolumn]{revtex4}
\pdfoutput=1
\usepackage{amssymb,latexsym}
\usepackage{amsmath,amsbsy,bbm}
\usepackage{ifpdf}
\usepackage{epsfig,bm}
\usepackage{graphicx,comment}
\usepackage{color}
\usepackage{soul}
\usepackage{mathtools}
\usepackage{comment}
\usepackage[normalem]{ulem}
\begin{document}

\title{The numerical value for a universal quantity of a two-dimensional
  dimerized quantum antiferromagnet}
\author{Fu-Jiun Jiang}
\email[]{fjjiang@ntnu.edu.tw}
\affiliation{Department of Physics, National Taiwan Normal University,
88, Sec.4, Ting-Chou Rd., Taipei 116, Taiwan}

\begin{abstract}
  The numerical value of a universal quantity associated with the quantum
  critical regime, namely $\chi_u c^2/T$, for a two-dimensional (2D) dimerized
  spin-1/2 antiferromagnet is
  calculated using the quantum Monte Carlo simulations (QMC).
  Here $\chi_u$, $c$, and $T$ are the uniform susceptibility, the spin-wave
  velocity, and the temperature, respectively. By simulating large lattices
  at moderately low temperatures, we find $\chi_u c^2/T \sim 0.32$. Our
  estimation of $\chi_u c^2/T$ deviates from the related analytic prediction
  but agrees with recent numerical calculations of other 2D dimerized
  spin-1/2 antiferromagnets.
  
\end{abstract}

\maketitle

\section{Introduction}\vskip-0.3cm

Spatial dimension two is special because of the famous Mermin--Wagner
theorem \cite{Mer66}. Specifically, continuous symmetry cannot be broken
spontaneously (at finite temperature $T$) in spatial
dimension two. As a result, for two-dimensional (2D)
quantum antiferromagnets, as $T$ rises from zero temperature, one
encounters a crossover instead of a phase transition to a unique phase
called the quantum critical regime (QCR).

Using the relevant field theory, properties of QCR are
investigated \cite{Chu94}. In particular, several universal quantities are
proposed. Among these quantities, we are especially interested
in $\chi_u c^2/T$ due to the recently found discrepancy between the analytic
prediction and the numerical calculations. Here
$\chi_u$ and $c$ are the uniform susceptibility and the spin-wave velocity,
respectively. 

Analytically, it is predicted that the numerical value of $\chi_u c^2/T$
is given by (around) 0.27185. Although this prediction was confirmed by earlier
Monte Carlo studies \cite{Chu94,Tro98}, recent investigations of
2D dimerized bilayer and plaquette spin-1/2 Heisenberg models
conclude that $\chi_u c^2/T \sim 0.32 (0.33)$ \cite{Sen15,Tan181}. Because of
this discrepancy, it will be
interesting to conduct a further examination on the numerical value of
$\chi_u c^2/T$.

In this study, we perform large-scale Monte Carlo simulations to determine
the $\chi_u c^2/T$ of a 2D dimerized quantum
antiferromagnetic Heisenberg model. By simulating lattices as large as
$L=512$ ($L$ is the linear system size), we obtain $\chi_u c^2/T \sim 0.32$
which matches quantitatively with recent outcomes claimed in
Refs. \cite{Sen15,Tan181}. Our result suggests that a refinement of analytic
calculation is needed.

The rest of the paper is organized as follows. After the introduction,
the model and the measured observables are described in Sec.~II.
We then present the obtained results in Sec.~III.
In particular, the numerical evidence to support $\chi_u c^2/T \sim 0.32$
is demonstrated. We conclude our investigation in Sec.~VI.

\begin{figure}
  \vskip-0.5cm
       \includegraphics[width=0.185\textwidth]{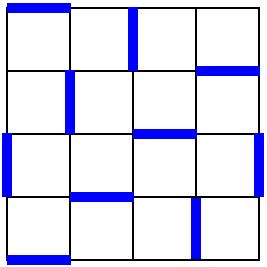}        
        \vskip-0.2cm
        \caption{The herringbone model considered in this study. The thick
          and thin bonds represent the coulplings of strength $J'$ and $J$,
          respectively}
        \label{fig0}
\end{figure}
  
\section{The considered model and observable}

The Hamiltonian of the considered 2D spin-$\frac{1}{2}$
dimerized herringbone Heisenberg model takes the following expression
\begin{eqnarray}
\label{hamilton}
H &=& \sum_{\langle ij \rangle}J\,\vec S_i \cdot \vec S_{j} 
+ \sum_{\langle i'j' \rangle}J'\,\vec S_{i'} \cdot \vec S_{j'}, 
\end{eqnarray}
where $J$ and $J'$ are the antiferromagnetic
couplings connecting nearest neighbor spins $\langle  ij \rangle$
and $\langle  i'j' \rangle$, respectively,
and $\vec{S}_i$ is the spin-$\frac{1}{2}$ operator at site $i$.
A cartoon representation of the studied model is depicted in 
fig.~\ref{fig0}.
$J$ is set to 1 in our investigation. As the magnitude of $J'$ increases,
a phase transition will occur for a particular value of $J' > J$.
This special point $J'/J$ in the parameter space is denoted by $(J'/J)_c$
and is found to be $(J'/J)_c = 2.4981(2)$ in the literature \cite{Pen20}.
The investigation presented in this study is conducted at the critical point
$(J'/J)_c$.

To examine the universal quantity $\chi_u c^2/T$ of QCR, the
uniform susceptibility $\chi_u$ and the
spin-wave velocity $c$ are measured.
On a finite lattice of linear size $L$, the uniform susceptibility $\chi_u$ is
defined by
\begin{equation}
\chi_u = \frac{\beta}{L^2}\left\langle\left(\sum_{i}S_i^z\right)^2 \right\rangle.
\end{equation}
The quantity $\beta$ appearing above is the inverse temperature.
The spin-wave velocity $c$ for the investigated model is
calculated through the temporal and spatial winding numbers squared
($\langle W_t^2\rangle$ and $\langle W_i^2 \rangle$ with $i\in\{1,2\}$).

\begin{figure}
  \vskip-0.5cm
  \includegraphics[width=0.325\textwidth]{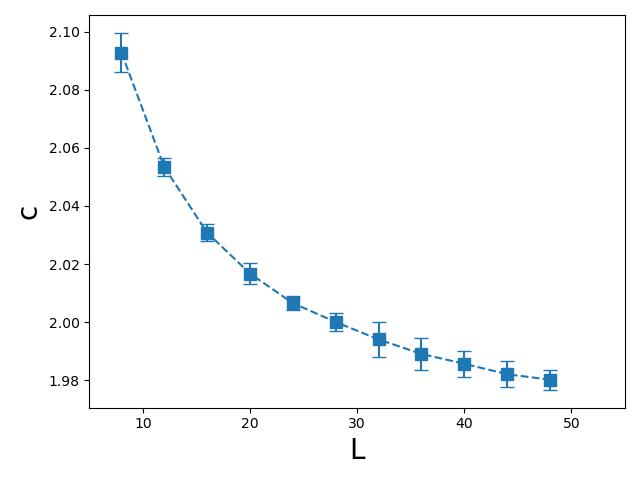}
        \vskip-0.2cm
        \caption{The spin-wave velocity $c$ as a function of the
          linear system size $L$.
        }
        \label{fig1}
\end{figure}

\begin{figure}
  \vskip-0.5cm
 \includegraphics[width=0.325\textwidth]{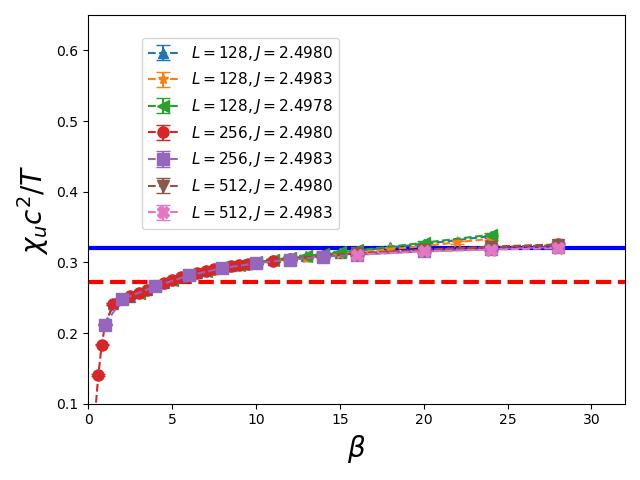}  
        \vskip-0.2cm
        \caption{
          $\chi_u c^2/T$ as functions
          of $\beta$
          for various values of linear system sizes $L$ and
          $\left(J'/J\right)_c$. The horizontal dashed and solid lines
        are 0.27185 and 0.32, respectively. }
        \label{fig2}
\end{figure}

\section{Numerical Results}

To conduct the proposed investigation, we have carried out large-scale
quantum Monte Carlo calculations (QMC) using the stochastic series
expansion algorithm (SSE) with efficent operate-loop update \cite{San99}.
The obtained outcomes are described in the following subsections.

\subsection{The determination of spin-wave velocity $c$}

The observable spin-wave velocity $c$ is required to compute the numerical value
of $\chi_u c^2/T$. Therefore we have calculated this
quantity first. 

We compute the spin-wave velocity $c$ 
by the method of winding numbers squared \cite{Sen15,Jia11}.
The procedure is as follows. For a given box size $L$ and a $J'/J$, 
the value of $\beta$ is tuned so that
$\langle W_t^2\rangle$ and $\langle W^2\rangle = \frac{1}{2}\sum_{i=1,2}\langle W_i^2\rangle$ match each other quantitatively. When this condition is fulfilled,
the spin-wave velocity $c(L,J'/J)$ associated
with this set of $L$ and $J'/J$ is given by $c(L,J'/J) = L/\beta$.

It should be pointed out that the existence of long-range antiferromagnetic
order is essential to employ this method to calculate $c$. As a result, we conduct the related simulations at $J'/J = 2.4975$ which is close to the critical point
$(J'/J)_c = 2.4981(2)$ and is in the antiferromagnetic phase.

The spin-wave velocity $c$ as a function of the linear system size $L$
is shown in fig.~\ref{fig1}.
The data in fig.~\ref{fig1}
is fitted to the equation of $a_0 + a_1/L + a_2/L^2 $. The determined $a_0$
is the bulk $c$ and is given by $1.952(5)$. The obtained
$c$ will be used in estimating the numerical value of $\chi_u c^2/T$.

\subsection{$\chi_u c^2/T$ as functions $\beta$}

The simulations of calculating $\chi_u$ are done with $L=128, 256$, and 512.
In addition, the uncertainty of $(J'/J)_c$ is taken into account in these
simulations. Using $c = 1.952(5)$, the $\chi_u c^2/T$ of various
$L$ and $(J'/J)_c$ as functions of $\beta$ are shown in
fig.~\ref{fig2}. The outcomes shown in the figure indicate that
as $L$ and $\beta$ increase,
the value of $\chi_c c^2/T$ deviates from its analytic prediction of 0.27185 (
the horizontal dashed line in the figure) and approaches 0.32 (the
horizontal solid line in the figure). This result agrees with that of
\cite{Sen15} and \cite{Tan181}.

\section{Discussions and Conclusions}

In this study, we calculate the numerical value of $\chi_u c^2/T$
corresponding to the 2D dimerized quantum herringbone Heisenberg model using
QMC. By simulating lattice as large as $L=512$ at moderate low
temperatures, we obtain $\chi_u c^2/T \sim 0.32$. Our result
deviates from the associated analytic prediction $\chi_u c^2/T = 0.27185$
but agrees quantitatively with recent numerical calculations
of 2D spin-1/2 bilayer and plaquette Heisenberg models. Based on
the obtained outcomes in this study, it will be interesting to
refine the relevant theoretical calculations. 


\section*{Acknowledgement}\vskip-0.3cm
Partial support from National Science and Technology Council (NSTC) of
Taiwan is acknowledged (MOST 110-2112-M-003-015 and MOST 111-2112-M-003-011).

\section*{Author Contributions}
F.J.J proposed the project, conducted the calculations, analyzed the data,
and wrote up the manuscript.

\section*{Conflict of Interest}
The author declares no conflict of interest.

\end{document}